\documentclass[11pt]{article}
\usepackage{mp,graphicx,epsfig}

\begin{document}

\begin{flushright}
TPJU-10/2002
\end{flushright}

\vspace*{3cm}
\title{PION GENERALIZED PARTON DISTRIBUTIONS \\IN THE NON-LOCAL NJL MODEL}

\author{\underline{M. PRASZA{\L}OWICZ} and A. ROSTWOROWSKI }

\address{M. Smoluchowski Inst. of Physics, Dept. of Theoretical Physics, Reymonta 4,\\
PL-30-059 Krak{\'o}w, Poland}

\maketitle\abstracts{ We calculate skewed parton distributions
$H^I$ of a pion within a non-local NJL model with momentum
dependent constituent quark mass $M(k)$. In the forward limit
$H^I$ correspond to parton distributions, whereas for $t<0$ $H^I$
is related to the pion electromagnetic form-factor}

\section{Pion in the-non local effective theory}

\label{subsec:prod}

Pion being a Goldstone boson of the broken chiral symmetry and a
$q\bar{q}$ bound state at same time has a unique status among
strongly interacting particles. Experiments that measure pion
generalized parton amplitudes (GDA) are therefore a direct tool to
investigate the dynamics of the chiral symmetry breaking. Yet,
compared to the nucleon, the experimental knowledge here is very
limited. The most precise are measurements of the pion form-factor
at low momentum transfer~\cite{piff}. There exist also data on
deep inelastic pion structure function~\cite{sf} measured in a
Drell-Yan process and prompt photon production. Both are related
to the pion skewed distribution~\cite{skewed}, $H^I$, at low
normalization scale. To calculate $H^I$ we use the non-local
chiral NJL model which follows from the instanton model of the QCD
vacuum~\cite{DPrev}. Here the external pion field
\begin{equation}
U^{\gamma _{5}}(x)=1+\frac{i}{F_{\pi }}\gamma ^{5}\tau ^{A}\pi ^{A}-\frac{1}{%
2F_{\pi }^{2}}\pi ^{A}\pi ^{A}+\ldots  \label{U}
\end{equation}
interacts with the constituent quarks
\begin{equation}
S_{I}=M \int \frac{d^{4}k}{(2\pi )^{4}}\bar{\psi}(k)F(k)
U^{\gamma_{5}}(k-l)F(l)\psi (l) .
\label{SI}
\end{equation}
$M \sim 350$~MeV is a constituent quark mass, $F_{\pi}=93$~MeV.
The cut-off function $F(k)$ is is known analytically only in the
Euclidean space. Here we wish perform calculations in the
Minkowski space. To this end we choose a simple pole formula
\cite{Bochum,Praszalowicz:2001wy}
\begin{equation}
F(k)=\left[ -\Lambda ^{2}/(k^{2}-\Lambda ^{2}+i\epsilon )\right]
^{n} \label{Fdef}
\end{equation}
We have already used this model \cite{Praszalowicz:2001wy} to
calculate pion light cone wave functions and various vacuum
condensates. Cutoff $\Lambda $ has been chosen in such a way that
the leading twist pion light cone wave function had proper
normalization.

$F(k)$ enters in pion-quark vertices as a form-factor which
provides an UV cutoff for the loop integrals and, secondly, it
also enters in the quark propagators as a momentum dependent mass.
In the present note we take into account both effects and in that
sense our calculations are exact. Technical details can be found
in Refs.[6].

\section{Skewed parton distributions in a pion}

Skewed parton distributions \cite{skewed} parameterize a
nonperturbative part of a deeply virtual Compton scattering (DVCS)
amplitude or hard meson production and are defined through the
matrix element \cite{skewed,WP}:
\begin{eqnarray}
& \int \frac{d\lambda }{4\pi }\exp (i\lambda X\,\,n\cdot P)\left\langle \pi
^{B}(p^{\prime })\right| \psi _{f}(-\frac{\lambda }{2}n)\rlap{/}n\,\psi _{g}(%
\frac{\lambda }{2}n)\left| \pi ^{A}(p)\right\rangle  \nonumber \\
& =\delta ^{AB}\delta _{fg}\,H^{I=0}(X,\xi ,t)+i\epsilon ^{ABC}\tau
_{fg}^{C}\,H^{I=1}(X,\xi ,t).  \label{Hdef}
\end{eqnarray}
Here $p$ ($p^{\prime }$) denote the incoming (outgoing) pion momenta, $%
P=(p+p^{\prime })/2$, $t=r^{2}=(p^{\prime }-p)^{2}$ and two null vectors $%
n^{\mu }=(1,0,0,-1)$, $\tilde{n}^{\mu }=(1,0,0,1)$. It is convenient to
choose a Lorentz frame in which (for $m_{\pi }=0$)
\begin{eqnarray}
P^{\mu } =\frac{1}{2}P^{+}\tilde{n}^{\mu }-\frac{t}{8P^{+}}n^{\mu
}, & & r^{\mu } =-\xi P^{+}\tilde{n}^{\mu }-\xi
\frac{t}{4P^{+}}n^{\mu }+r_{T}^{\mu }.  \label{kin}
\end{eqnarray}
Note that the Fourier transform in (\ref{Hdef}%
) introduces a kinematical parameter $X$ which in the forward limit ($%
t=0 $ and $\xi =0$) equals to the momentum fraction
$X=k^{+}/P^{+}$ carried by a struck quark.

In the forward limit functions $H^{I}(X,0,0)$ defined in
(\ref{Hdef}) are related to the parton densities $q(X)$. For $\pi
^{+}$ \emph{e.g.} we have
\begin{equation}
2H^{I=0}(X)=\left\{
\begin{array}{c}
u(X)+\bar{u}(X)\\ 
-\bar{d}(-X)-d(-X) 
\end{array}
\right.   \, \,
H^{I=1}(X)=\left\{
\begin{array}{ccc}
u(X)-\bar{u}(X) & \rm{for} & X\geq 0 \\
\bar{d}(-X)-d(-X) & \rm{for} & X\leq 0
\end{array}
\right. .  \label{q2}
\end{equation}

For $t<0$
\begin{equation}
\int_{-1}^{1} dX\,H^{I=1}(X,\xi ,t)=F_{\pi }^{\rm{em}}(t).
\label{ff}
\end{equation}
which is true for any $\xi $. In what follows we shall use
relations (\ref{q2}) and (\ref{ff}) to extract both pion
form-factor and quark densities.

\begin{figure}[t]
\centerline{
\includegraphics[width=.28\textwidth]{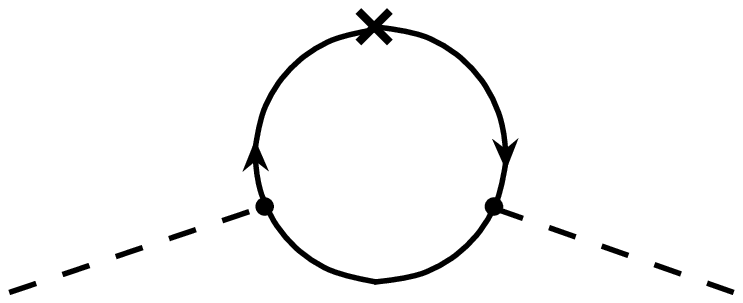}
\includegraphics[width=.28\textwidth]{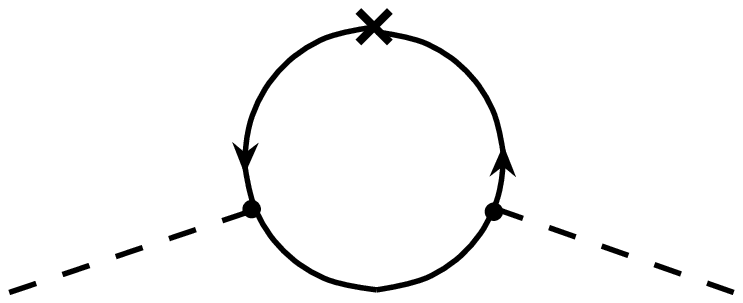}
\includegraphics[width=.28\textwidth]{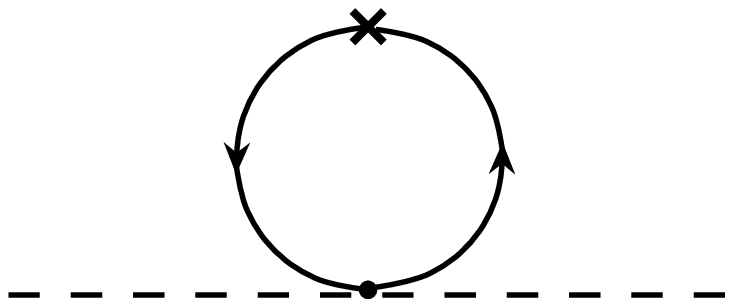}}
\caption{Graphical representation of Eq.(8). Cross denotes
external current, dashed line external pion fields.
\label{fig:loop}}
\end{figure}

\section{Numerical Results}

The contribution to $H^{I}$ is given in terms of 3 Feynman
diagrams shown in Fig.1:
\begin{eqnarray}
H^{I=0} =\mathcal{I}_{1}+\mathcal{I}_{2}+\mathcal{I}_{3}, & &
H^{I=1} =\mathcal{I}_{1}-\mathcal{I}_{2}.
\end{eqnarray}

In Fig.2a. where we plot $H^{I=0}(X,\xi =0.5,t=-0.25$
Gev$^{2}/c^{2})$ the dashed-dotted (black) line corresponds to the
sum $\mathcal{I}_{1}+\mathcal{I}_{2}$, whereas the dashed (blue)
line corresponds to $\mathcal{I}_{3}$ which is non-zero only for
$-\xi <X<\xi $. The sum is depicted by the solid (green) line.
These results agree with the ones  of Refs.[7]\nocite{WP}. In
Fig.2b. we plot $2H^{I=0}(X,\xi =0,t=0$ GeV$^{2}/c^{2})$.  Since
in our model the pion ($\pi ^{+}$ for definiteness) is build only
from valence quarks, the right part ($X>0)$ corresponds to
$u_{v}(X)$, whereas the left part ($X<0$) describes
$-\bar{d}_{v}(-X)$. We see very weak dependence on
the parameter $n$  entering Eq.(\ref{Fdef}). In Fig.2b we also plot the $%
H^{I=0}$ for a constant constituent mass $M$ (dashed (green) line), where
the pertinent loop integrals were regularized by a sharp cutoff in the
transverse momentum plane. This form of the valence quark distributions $%
u_{v}(X)=\bar{d}_{v}(X)=\Theta (X(1-X))$ have been recently
advocated~\cite{DRA} as the one compatible with the Ward
identities.

\begin{figure}[h]
\centerline{\includegraphics[width=.5\textwidth]{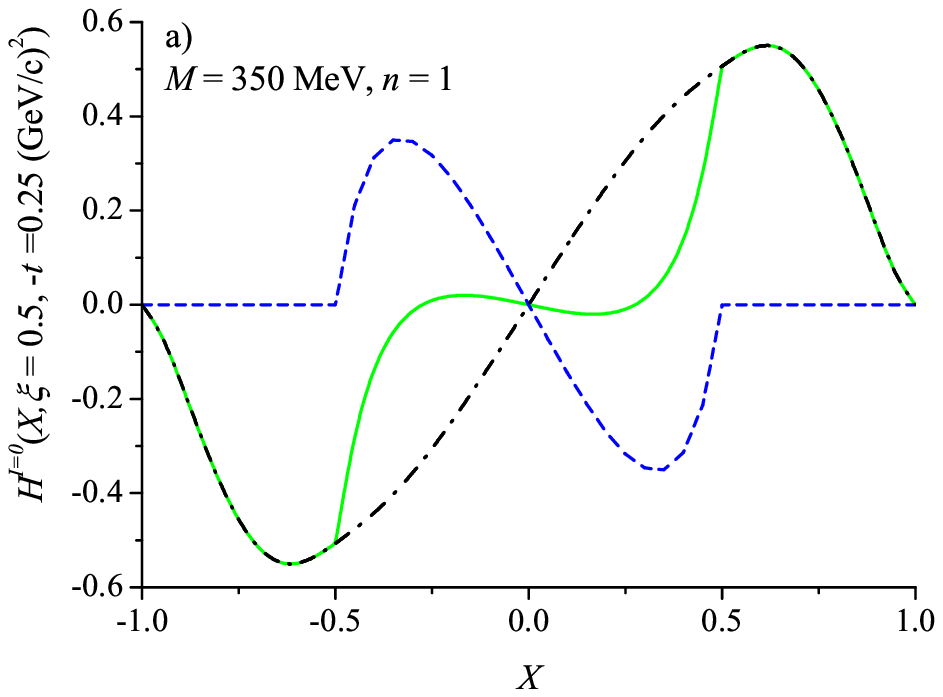}
\includegraphics[width=.5\textwidth]{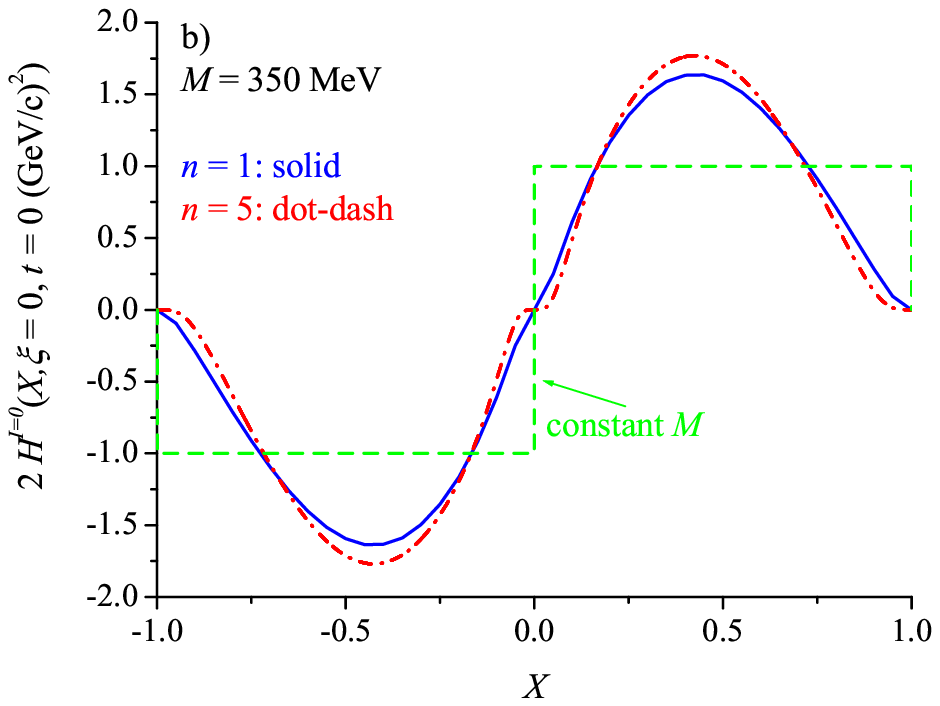}}
\vspace{-5pt} \caption{Skewed pion distribution $H^{I=0}$ for
$t<0$ and in the forward limit. \label{fig:H}}
\end{figure}

The quark distributions calculated in this way correspond to some
low normalization scale $Q_{0}$ which is of the order of a few
hundred MeV~\footnote{In the instanton model this scale is
associated with the inverse of the average instanton size $1/\rho
=600$ MeV .}. In order to make contact with the existing analysis
of the pion nucleon data~\cite{sf} at $Q=2$ GeV, we have evolved
the distributions depicted in Fig.2b starting from two different
scales $Q_{0}=450$ and $350$ MeV. The results are plotted in
Fig.3a together with the two parameterizations of the experimental
data~\cite{sf}: GRS  and SMRS. The dashed (green, DR-A) line
corresponds to the constant $M.$ Surprisingly the latter fits the
experimental analysis the best \cite{DRA}. It would be of interest
to calculate the Drell-Yan cross section directly using the model
qurak distributions and compare with the data.

Finally in Fig.3b we show the pion electromagnetic form-factor
calculated by means of Eq.(\ref{ff}), which -- as we checked
explicitly -- does not depend on $\xi $. One can see from Fig.3b
that the theoretical curve overshoots the experimental points.

\begin{figure}[t]
\centerline{\includegraphics[width=.5\textwidth]{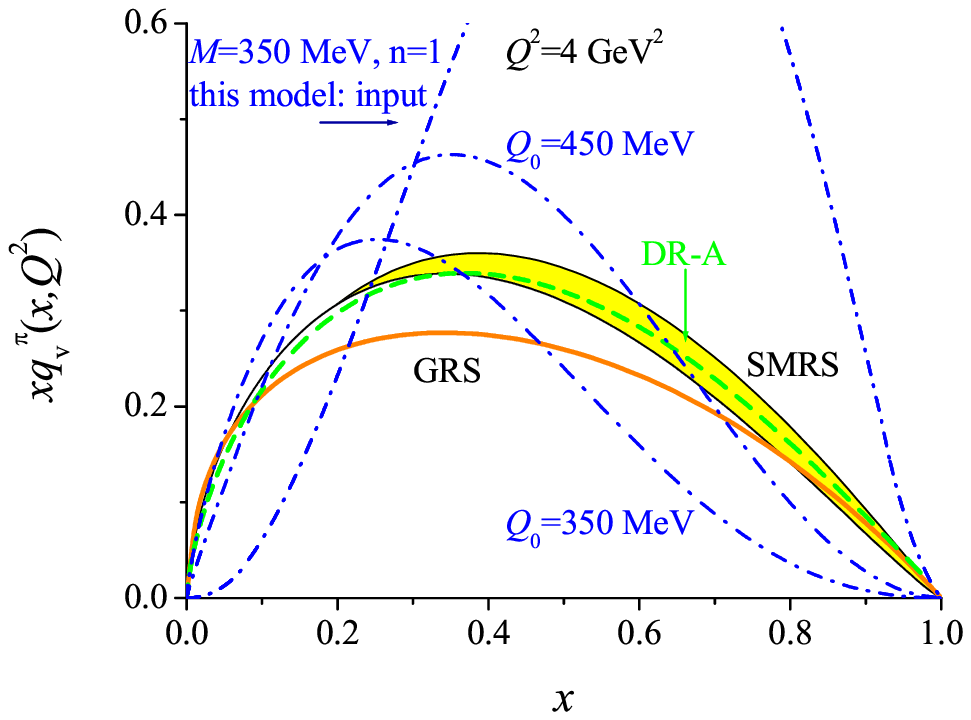}
\includegraphics[width=.5\textwidth]{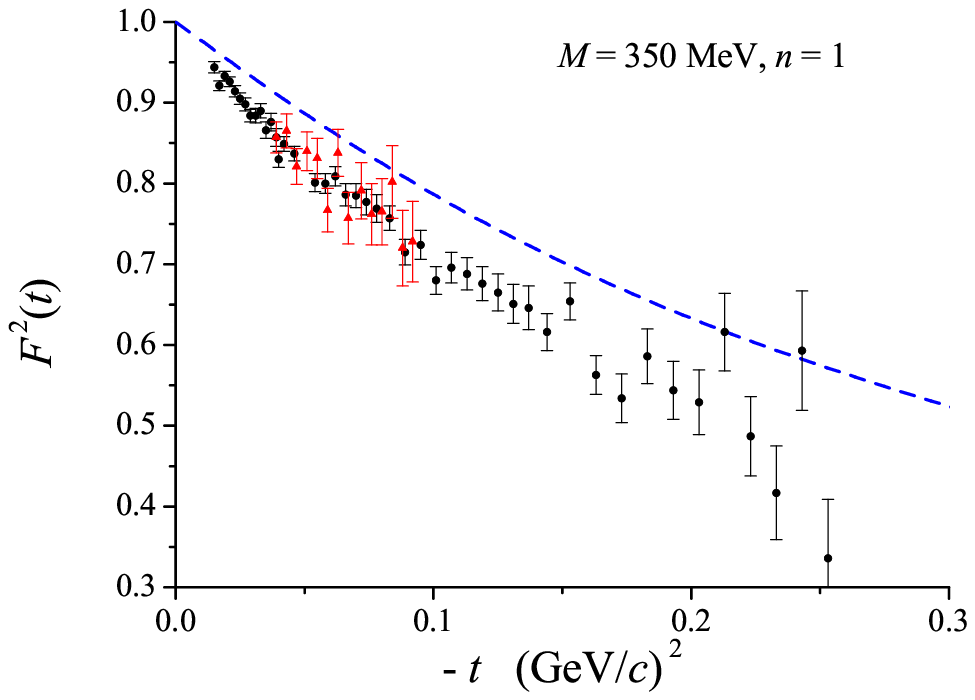}}
\vspace{-5pt}
\caption{Quark distributions and pion
electromagnetic form-factor \label{fig:sfff}}
\end{figure}

\section{Concluding remarks}

In calculating the pion skewed distributions we have relied on the
definition (\ref{Hdef}) without taking into account that the
currents in the non-local theories acquire additional terms
\cite{nonl}. It is of importance to investigate the impact of
these new terms on our results. Pion structure function has been
calculated in a similar model but in a different way in
Ref.[10]\nocite{Toki}. An alternative approach consists in
calculating entire amplitude in the effective model \cite{AD}
without any reference to the factorization theorems.

\section*{Acknowledgments}

M.P. wishes to thank the organizers of the Moriond 2002 for an
invitation to this very stimulating meeting. We would like to
thank W. Broniowski, A.E. Dorokhov, M.V. Polyakov, E. Ruiz-Arriola
for discussions. Special thanks are due to K. Golec-Biernat for
making the DGLAP evolution code available to us.  This work was
partially supported by Polish KBN Grant PB~2~P03B~{\-}019~17 and 
Polish KBN Grant 2 P03B 048 22.

\end{document}